\documentstyle[aps,preprint,epsf]{revtex}

\begin{document}

\def\d{\delta}
\def\e{\epsilon}
\def\s{\sigma}
\def\l{\lambda}
\def\.{\cdot}
\def\o{\omega}
\def\a{\alpha}
\def\b{\beta}
\def\be{\begin{eqnarray}}
\def\nn{\nonumber}
\def\ee{\end{eqnarray}}
\def\bi{\begin{itemize}}
\def\i{\item}
\def\ei{\end{itemize}}

\def\({\left(}
\def\[{\left[}
\def\){\right)}
\def\]{\right]}
\def\h{{1\over 2}}
\def\eq#1{(\ref{#1})}
\def\tr{{\rm Tr}}
\def\bkk#1{\langle#1\rangle}
\def\bk{{\bf k}}
\def\bx{{\bf x}}
\def\bd{{\bf\Delta}}
\def\bb{{\bf b}}
\def\bp{{\bf p}}
\def\bq{{\bf q}}
\def\bz{{\bf 0}}
\def\b1{{\bf 1}}
\def\BG{{\bf \overline G}}
\def\A{{\cal A}}
\def\x{\times}
\def\g1{\b1\!\x\!\b1}
\def\gc{{\bf t'_a}\times{\bf t''_a}}
\def\D{\Delta}
\def\C{{\bf C}}
\def\G{{\bf G}}
\def\sa{\sim\hskip-3pt\alpha_s}

\def\labels#1{\label{#1}}

\title{Unitarized Diffractive Scattering in QCD\\ and Application to Virtual
Photon Total Cross Sections}
\author{Rim Dib, Justin Khoury, and C.S. Lam}
\address{Department of Physics, McGill University,
3600 University St., Montreal, QC, Canada H3A 2T8\\
E-mails: rdib@po-box.mcgill.ca, 
jkhoury@princeton.edu, Lam@physics.mcgill.ca}
%\twocolumn

\maketitle

\begin{abstract} The problem of restoring Froissart bound to the
 BFKL-Pomeron is studied in an extended leading-log approximation of QCD. 
We consider parton-parton scattering amplitude and 
show that the sum of all Feynman-diagram contributions can be written 
in an eikonal form. In this form dynamics is determined by
the phase shift, and subleading-logs of all orders needed to restore the
Froissart bound are automatically provided.
The main technical difficulty
is to find a way to extract these subleading contributions without having to 
compute each Feynman diagram beyond the leading order. We solve that problem
by using nonabelian cut diagrams introduced elsewhere. 
They can be considered as
colour filters used to isolate the multi-Reggeon 
contributions that supply these subleading-log terms. 
Illustration of the formalism is given for amplitudes and phase shifts up to 
three loops. For diffractive scattering,
only phase shifts governed by one and two Reggeon
exchanges are needed. They can be computed from the leading-log-Reggeon 
and the BFKL-Pomeron amplitudes. In applications,
we argue that the dependence of the energy-growth
exponent on virtuality $Q^2$ for $\gamma^*P$ total cross section observed
at HERA can be interpreted as the first sign of a slowdown of energy
growth towards satisfying the Froissart bound. An attempt to understand these 
exponents with the present formalism is discussed.
\end{abstract}

\section{Introduction}
Large rapidity-gap events observed recently at HERA \cite{HERA,ZEUS}
provide additional impetus to QCD calculation of diffractive 
scattering and total cross section. At virtuality $Q^2\gg\Lambda_{QCD}^2$,
the QCD fine structure constant $\alpha_s=g^2/4\pi$ is small, near-forward
parton-parton scattering is in principle calculable in perturbation
theory, and total cross section can  
be obtained via the optical theorem.
To be sure, such calculation 
is very complicated at high c.m. energy $\sqrt{s}$, 
as the  effective expansion parameter for the problem
is $\alpha_s\ln s$ and not just $\alpha_s$. 
The former could be large at high $s$ even though 
the latter may be small. As a result, multi-loop diagrams must be included.

To make multi-loop calculations feasible,
{\it leading-log approximation} (LLA) is usually employed. This
consists of keeping only the highest power of $\ln s$ at each order. Equivalently, if the scattering amplitude (divided by $s$)
is considered as a function of the two variables $\alpha_s$
and $\alpha_s\ln s$, then only the
lowest power of $\alpha_s$ is kept.
When the resulting amplitude is of the form 
$\alpha_s^nsF(\alpha_s\ln s)$
for some function $F$, we will abbreviate it simply as $\sa^n$.

The LLA result for high-energy 
near-forward scattering is known, though not completely.
 The dominant parton-parton scattering
amplitude occurs with the exchange of a colour-octet
object called the Reggeon (or the Reggeized gluon), and is $\sa$ \cite{DEL}. 
The
Reggeon amplitude is actually proportional to
$(\alpha_ss/\Delta^2)s^{R(\D)}$, with $\Delta^2=-t$
the square of the momentum transfer, and $R(\Delta)$ 
a known function of $\Delta$ proportional to $\alpha_s$. Note that
$s^{R(\Delta)}$ is of the form $F(\alpha_s\ln s)$, so the Reggeon amplitude
is indeed $\sa$, as claimed. Although the Reggeon amplitude is
the dominant result for summing {\it all} diagrams in LLA, 
it can be obtained
just from the $t$-channel ladder diagrams and their gauge partners,
because all the other diagrams are subdominant.
We may therefore think of the Reggeon as a composite object made up of gluons, bundled up together roughly in a ladder configuration.

Diffractive scattering occurs via a colour-singlet exchange. 
Two or more Reggeons are required to form a colour-singlet object
in the $t$ channel. Since each Reggeon carries a small factor $\sa$, 
the dominant singlet amplitude comes from two interacting Reggeons, giving
a magnitude $\sa^2$.
This is the BFKL Pomeron \cite{DEL,BFKL}.
Unfortunately this amplitude is not known too precisely, 
though one does know that
in the forward direction and extremely high energies it behaves like
$s^J$, with $J>1$  a known number. 
This gives rise to a total cross section with energy variation
$s^{J-1}$, which violates the Froissart bound $(\ln s)^2$.

Present data from HERA \cite{HERA,ZEUS} on $\gamma^*P$ 
total cross section are 
consistent with a power growth in $s$, though with an observed exponent
smaller than $J-1$ computed from the BFKL Pomeron. 
To satisfy unitarity and obey the Froissart bound, the exponent must 
eventually decrease to zero at very high energies,
and that can be achieved 
theoretically only by including subleading-log contributions that have
been hitherto neglected.
The first subleading correction
to the BFKL Pomeron have been computed \cite{BFKL2}, 
with the encouraging result of a smaller $J$, but also with
a pathological behaviour that has
not yet been fully resolved. 
In any case the Froissart bound is
still violated.

From the $s$-channel unitarity relation
2Im$(T_{fi})=\sum_nT^*_{fn}T_{ni}$, one expects all multiple-Reggeon
exchanges to have to be included before we can restore full unitarity
and the Froissart bound.
As each additional Reggeon brings in an extra factor of $\sa$, 
they provide the required subleading-log
contributions to all orders. The purpose of this paper is to discuss
a formalism whereby this scenario can be implemented in pQCD.
For other approaches see for example \cite{SAT}.

To clarify what we have in mind, let us first pretend the
partons to be scattered 
through a two-body instantaneous potential $V(\vec x)$.
Born amplitude alone violates unitarity, but that can be fixed up
by including all higher-order corrections. For small $V$, these higher-order
terms are the analogs of the subleading contributions discussed above.
These amplitudes can be summed up at high energy into 
an eikonal formula  (see eq.~\eq{impact}) \cite{SAKURAI}
where unitarity is restored. Dynamics is now specified by
the phase shift $\delta(b)$, which is a function of the impact
parameter $b$, and is linear in $V$.
Higher-order amplitudes
can be recovered by expanding the exponential
into powers of the phase shift. 
 
A very similar scenario emerged if the two partons interacted through a 
multiple exchange of photons. In that case the interaction is no longer 
instantaneous, and cross
diagrams are present. In terms of the unitarity relation, 
this means photons
are present in the intermediate states and the inelastic channels 
in which they are produced must be included in the unitarity sum.
However, the eikonal form can still be established with the help
of the `eikonal formula' \cite{EIK}, 
and the phase shift is still given by the Born approximation amplitude 
(the Coulomb phase shift) \cite{STERMAN,CW}. 

There is nothing new about using the  eikonal form
to implement unitarity. In fact, a very large amount
of phenomenology of two-body scattering has been carried out in that
framework. The modern challenge is whether, and if so how, 
pQCD amplitudes can be unitarized that way. 
For on-shell near-forward
hadronic scatterings, confinement is presumably important
and we may not even be able to use pQCD. However,
$\gamma^*P$ total cross section at
large virtuality is expected to be calculable in pQCD, and
the BFKL Pomeron is an attempt to do so. Unfortunately
it violates the Froissart bound, so the implementation of unitarity in
pQCD remains unsolved. This is the area where we hope to make
some progress on. The main difficulty comes from the fact that
terms with all powers of $\alpha_s$ (with fixed $\alpha_s\ln s$) 
are needed to restore 
unitarity, but these are small terms extremely difficult to obtain 
from Feynman diagrams in the calculable regime $\alpha_s\ll 1$.

The following physical picture \cite{GRIBOV} may help to
visualize how the Froissart bound is restored.
The rise of total cross section \cite{CW0} predicted by the BFKL
Pomeron \cite{BFKL} may be attributed to an increased  production
of gluon jets at higher energies. When the energy gets really high,
there are so many gluons around that they tend to overlap one another.
When that happens, coherent effect becomes important and a 
destructive interference sets in to reduce the power growth, 
to a rate that eventually satisfies
the  Froissart bound $(\ln s)^2$.
This mechanism suggests that whatever formalism we use to restore
the Froissart bound, interference effects ought to be a central part of
it. The `nonabelian cut diagrams' we propose to use to solve this problem
fits into this category, because they can be viewed as a way 
to organize the summation of Feynman diagrams
to heighten the Bose-Einstein interference effects of 
identical gluons \cite{LCONF1}.

Returning to the formal mechanism to unitarize QCD, 
the first thought would be to
try to imitate potential scattering and QED
by including multiple exchanges of Reggeons and/or Pomerons.
This is indeed the general idea but there are 
non-trivial problems to be solved with this approach.
Unlike the photon
in QED or the potential, Reggeon and Pomeron are composite objects, themselves made up of gluons and possibly quark pairs. 
It is therefore not clear whether we
are allowed to exchange them as if they were elementary. 
For example, should we include diagrams where Reggeons are crossed?
Or equivalently, can Reggeons be produced from the partons so that 
inelastic channels involving their production must be included in the
unitarity sum like QED? Can these composite objects overlap
and merge? If an 
effective theory of Reggeon/Pomeron equivalent to QCD existed
without gluons, then presumably we would be able to answer all
these questions once its precise dynamics is known. 
However, since gluons can be produced off Reggeons and Pomerons, they
must be included in any effective theory \cite{EFFLAG}, in which case 
it is hard to exclude them from being exchanged as well. 
If we have to exchange gluons anyhow, we might
as well go back to the original QCD theory which exchanges nothing
but gluons and quarks, and whose dynamics is precisely known. At least we can
avoid potential double counting if we do it that way.
Nevertheless, gluon self-interaction and non-commutativity of colour matrices
have no analog in the potential or the QED problem, so 
it is not clear how the very complicated QCD
Feynman diagrams can be handled.
It is even less certain that the results can be mimicked by the exchange
of Reggeons and/or Pomerons which we hope to see.
Worst of all, Feynman diagrams can hardly be calculated beyond LLA, but
we need subleading contributions to build up the eikonal form and unitarity.
How can we possibly get them without going through the impossible
task of computing Feynman diagrams to highly subleading orders?  
   
Fortunately these questions can be answered and difficulties overcome,
if we use the nonabelian
cut diagrams introduced elsewhere \cite{LL1,LJMP,FHL,FL} instead
of the conventional Feynman diagrams. 
Reggeons and Pomerons emerge naturally, and phase shifts can be
calculated using only the  leading-log approximation.
Due to the possibility of gluon production from Reggeons,
phase shift is generally no longer linear in the (Reggeon)
exchange, as was the case in potential scattering and QED.

Nonabelian cut diagrams will be reviewed in Secs.~3 and 4. 
They are Feynman diagrams with slightly different `Feynman rules'.
A permuted sum of Feynman diagrams can be shown to be equal to
a permuted sum of (nonabelian) cut diagrams, which is why we may
calculate amplitudes using either of them.
The main advantage of cut diagram is that
identification with Reggeon becomes natural. 
In fact, each cut amplitude is given by a product of Reggeon (fragment)
amplitudes. This factorization allows the 
eikonal form to be built up, and subleading terms to be computed 
just in the leading-log approximations.

We have used the word {\it Reggeon fragment amplitudes} 
to mean finite-order amplitudes whose sum builds up the full Reggeon amplitude.
Hence the word `fragment'.
There may be many distinct fragments even at a given order.
For brevity, the word `fragment' will often be dropped.

In Sec.~2, the impact-parameter representation of a high-energy two-body
amplitude is reviewed. A physical argument is given to show how the
Froissart bound is restored when the bound-violating Born approximation
is iterated and summed up into an eikonal form. 
In Sec.~3, nonabelian factorization formula 
\cite{LL1,LJMP} and
nonabelian cut diagrams \cite{FHL,FL} for tree amplitudes are reviewed;
nonabelian cut diagrams are simply Feynman diagrams with the factorization
formula built in. It is important to note that this factorization
has nothing to do with the usual factorization of hard physics from 
soft physics \cite{STERMAN}. The present one goes along the $s$-channel,
whereas the usual one goes along the $t$-channel.
In Sec.~4, the technique is applied to two-body scattering amplitudes.
The advantage of using cut diagram over Feynman diagram
to compute a permuted sum is discussed. The connection with Reggeons
is identified.
The question of factorizability of these
amplitudes into irreducible parts is discussed in Sec.~5. 
This factorization allows the eikonal form to be built
up and phase shifts computed from the irreducible amplitudes.
Diffractive scattering is 
considered in Sec.~6. In that case phase shift may be restricted 
to those obtained from one and two irreducible Reggeon exchanges, and
can be computed from the 
LLA-Reggeon and BFKL-Pomeron amplitudes. By using phase
shifts in the eikonal form unitarization of the diffractive
amplitude is achieved.
To illustrate the details of previous sections,
quark-quark scattering amplitudes and phase shifts are given 
to the three-loop order in Sec.~7.
Sec.~8 is devoted to the observed $\gamma^*P$ total cross sections.
We argue that the dependence of the energy-growth exponent on
virtuality $Q^2$, as well as the size of the BFKL exponent compared
to the observed ones, can be taken as indications that the softening
of energy growth towards the Froissart-bound limit is already happening.
Correct energy dependence at various
virtuality $Q^2$ can be reproduced with the formulae of Sec.~7.
Finally, Appendix A contains some colour-algebra calculations.

\section{Impact-Parameter Representation, Phase Shifts, \\ and the Froissart
Bound}
Let $A(s,\Delta)$ be a parton-parton scattering amplitude at 
centre-of-mass energy $\sqrt{s}$
and momentum transfer $\Delta=|\bd|$. At high energies
$\bd$ is transverse, and conjugate to the impact-parameter $\bb$.
Bold letters like $\bd$ and $\bb$ are used to describe vectors
in the transverse plane.

The two-dimensional Fourier transform of $A(s,\Delta)$ defines
the impact-parameter amplitude $\A(s,b)$ by the eikonal
formula
\be
A(s,\Delta)=2is\int d^2\bb e^{i\bd\.\bb}\A(s,b)\equiv
2is\int d^2\bb e^{i\bd\.\bb}\(1-e^{2i\delta(s,b)}\).
\labels{impact}\ee
For large $s\simeq 4k^2$, angular momentum is given by $l=kb$,
and $\delta(s,b)$ is just the phase shift at that angular
momentum. For QCD parton scatterings, the initial and final partons
may contain different colours, 
so the amplitudes $A(s,\Delta)$ and $\A(s,b)$,
as well as the phase shift $\d(s,b)$, should be treated as colour matrices.
Only the diagonal matrix elements arising from
a colour-singlet exchange are truly elastic amplitudes. We shall
denote them by $A_1(s,\Delta)$ and $\A_1(s,b)$. In terms of them,
the total cross section
$\s_T(s)$ is given by the optical theorem to be
\be
\sigma_T(s)={1\over s}{\rm Im}\[A_1(s,0)\].
\labels{tot}
\ee

No approximation has been made to arrive at \eq{impact},  
so unitarity is
exact and the Froissart bound is satisfied. If $\delta(s,b)$
is small, the exponential may be expanded and 
only the term
linear in $\d(s,b)$  kept. This gives the 
`Born amplitude' and allows the phase shift to be thought of 
as the `potential'.
Thus if the interaction causing the scattering has a 
range $\mu^{-1}$, then we expect
$\delta(s,b)\sim\! a\exp(-\mu b)$. 
At large impact parameters,
$a$ may be taken to be approximately independent of $b$. 
In the region $\mu b\gg 1$,
the phase shift is small and the impact amplitude $\A(s,b)$ in \eq{impact}
ceases to contribute. The
effective radius $R(s)$ of interaction may therefore
be estimated from the condition
$\delta(s,R)\sim 1$. If $a$ is an increasing function
of $s$, then $R(s)$ and hence $\sigma_T\sim \pi R(s)^2$ 
also increase with $s$. In particular, if $a\sim s^{J-1}$ has a power 
growth, caused for example by the exchange of a spin $J>1$ Pomeron
 or elementary particle,
 then $R(s)\sim [(J-1)/\mu]\ln s$, and $\sigma_T\sim (\ln s)^2$,
which is the Froissart bound. 
This conclusion is 
general and is
qualitatively independent of the magnitude of $J$
and the range $\mu^{-1}$. Unless 
$a$ increases faster than a power of $s$, the
 Froissart bound is  guaranteed when
total cross-sections are calculated from the phase shift.
On the other hand, if we used the `Born approximation' throughout,
then the total cross section would have
grown like $s^{J-1}$ all the way, and the Froissart bound
would be violated. This is
essentially what happens to the BFKL Pomeron \cite{BFKL}. 
The cure, very roughly speaking,
is to use it as a phase shift rather than an amplitude. 
This is the general idea but
details are more complicated. They will be discussed in
Sec.~6.

To make use of this unitarization mechanism  we must 
find a way to calculate the phase shift. This is not
simple for several reasons.
According to \eq{impact},
even when $\delta(s,b)$ is computed just to
the lowest order, the resulting $A(s,\Delta)$ contains terms of
all orders. This suggests that a proper understanding of phase
shifts cannot be obtained until we know how to sum
an infinite number of Feynman diagrams. To put it
differently, we have to learn how to deal with
$2i\d(s,b)=\ln(1-\A(s,b))$, which consists of an infinite sum
of all powers of the matrix amplitude $\A(s,b)$.

Moreover, according to \eq{impact}, to be successful a certain product
structure must emerge out of the sum.
If the phase shift is expanded in powers of the coupling constant
$g^2$, $\delta(s,b)\simeq \sum_mg^{2m}\delta^{(m)}$, then 
the $(2n)$th order contribution to $\A(s,b)$ is given by a sum of products
of the phase shifts $\delta^{(m_i)}$, with $\sum m_i=n$. 
Individual Feynman diagrams certainly
do not factorize in this manner, and it is not immediately clear why
sums of Feynman diagrams have this structure either. But unless we can
get the sum into this factorized form there would seem to be no simple way to extract the phase shift from Feynman diagrams.

In the next two sections we shall lay down the foundation 
which enables us to build up such a factorized form.

\nopagebreak
\section{Eikonal Approximation and Factorization 
of Nonabelian Tree Amplitudes}
The decomposition of sums of Feynman diagrams into 
sums of products of Reggeon amplitudes, or of phase shifts,
 is based on a nonabelian
{\it factorization formula} \cite{LL1,LJMP}. This formula
for tree amplitudes is the nonabelian generalization of the 
{\it eikonal formula} \cite{EIK} used
in high-energy scattering in QED. It can be conveniently
embedded into Feynman diagrams to turn them into {\it nonabelian cut
diagrams} \cite{FHL,FL}. These items will be
reviewed  in the present section.

The formula deals with a sum of tree amplitudes like
Fig.~1, in which a
fast particle of final
momentum ${p'}^\mu$ emerges from an initial particle with momentum $p^\mu$
after the emission of 
$n$ gluons of momenta $k_i^\mu$.
Spins and vertex factors 
are ignored in this section but they will be incorporated
later. The gluons concerned
need not be on shell. This allows the tree diagram to be 
a part of a much larger Feynman diagram, so that formulae developed
for trees here are also useful for more complicated diagrams later.
We shall assume the energy $p^0\simeq {p'}^0$ of the fast particle 
to be much larger
than its mass $m$, its transverse momentum $\Delta=|\bd|$,
the square-root virtuality $Q$ in case it is present, and all the components
$k_i^\mu$ of every gluon momentum; in short, larger than any other
energy scale involved. It is convenient to use a somewhat unconventional
system in which the final particle moves along the z-axis. 
In the light-cone coordinates defined by $A^\pm=(A^0\pm A^3)$,
the final four-momentum of the energetic particle 
can then be written as ${p'}^\mu=(p^+,p^-,\bp)=(\sqrt{s},0,0)$, where terms
of $o(1)$ have been dropped. The initial momentum $p^\mu$ of the fast
particle then carries a transverse component $\bd=\sum_{i=1}^n\bk_i$.
In the  {\it eikonal approximation} outlined above, 
\be
{1\over (p'+K)^2-m^2+i\e}={1\over 2p'\.K+K^2+i\e}
\simeq{1\over 2p'\.K+i\e}\simeq{1\over \sqrt{s}}{1\over K^-+i\e}
\labels{eik}
\ee
can be made on all propagators of the fast particle, provided 
$K$ is the sum
of any number of the $k_i$'s. 
The factor $s^{-\h}$ in \eq{eik} is irrelevant for the
rest of this section so it will be dropped and  the 
propagators taken simply to be $(K^-+i\e)^{-1}$.

The tree amplitude in Fig.~1 is then given by the product of a momentum
factor $a[\tilde\s]$ and a colour factor $t[\tilde\s]$, where
\be
a[\tilde\s]&=&-2\pi i\d\(\sum_{j=1}^nk^-_j\)
\prod_{i=1}^{n-1}{1\over\sum_{j=1}^ik^-_i+i\e},\nn\\
t[\tilde\s]&=&t_1t_2t_3\cdots t_n.
\labels{tree}
\ee
The $\d$-function in $a[\tilde\s]$ is there to ensure the 
initial-state momentum $p=p'+\sum_{i=1}^nk$ to be on shell, 
for in the eikonal 
approximation its square is  $m^2+\sqrt{s}\sum_{i=1}^n
k^-_i$. Since this $\d$-function is not explicitly 
contained in the T-matrix,
it should be removed at the end, but for the sake of a simple statement 
in the
factorization formula \eq{fact} below it is convenient to include it
in $a[\tilde\s]$ for now.

The colour factor $t[\tilde\s]$ is given by a product of
$SU(N_c)$ colour matrices $t_a$
 in the representation appropriate to the fast particle.

Since gluons obey Bose-Einstein statistics, we must sum over all 
their permutations
to obtain the complete tree amplitude. Let $[\s]=[\s_1\s_2
\cdots\s_n]$ indicate the ordering of 
gluons along the fast particle from left to right. This
symbol will also be used to denote the
corresponding tree diagram.
In this notation Fig.~1 corresponds to $[\s]=[\tilde\s]\equiv[123\cdots n]$.
The tree amplitude for the diagram $[\s]$ will be denoted by
$a[\s]t[\s]$;
they are given by \eq{tree} with appropriate permutation of the 
gluon momenta.
The complete tree amplitude is given by the sum of individual
tree amplitudes
over the $n!$ permutations $[\s]$ of the permutation group $S_n$.
The {\it factorization formula} \cite{LL1,LJMP} states 
that this permuted sum can be replaced by a similar sum of the
{\it nonabelian cut amplitudes}:
\be
\sum_{[\s]\in S_n}a[\s]t[\s]=\sum_{[\s]\in S_n}a[\s_c]t[\s_c'].
\labels{fact}
\ee 
Just like the amplitudes on the left-hand side which
can be obtained from Feynman diagrams, 
the (nonabelian) cut amplitudes $a[\s_c]t[\s_c']$ 
on the right-hand side can be obtained from {\it nonabelian cut diagrams}\cite{FHL,FL}. 
The (nonabelian) cut diagrams are 
Feynman diagrams with cuts inserted at the appropriate
 propagators. The `Feynman rules' for cut diagrams
are the usual ones except at and around the cut propagators, as we
shall explain.
The position
of the cuts depends on $[\s]$, and is given by the following rule. 
{\it A cut is placed just to the right of
gluon $\s_i$ if and only if $\s_i<\s_j$ for all $j>i$}. We shall
label a cut by a vertical bar, either on the diagram itself or in
the corresponding permutation symbol $[\s]$. 
The resulting nonabelian cut diagram will be denoted by $[\s_c]$. 
For example, if $[\s]=[32145]$, then $[\s_c]=[321|4|5]$. If
$[\s]=[12354]$, then $[\s_c]=[1|2|3|54]$.

The momentum factor of a cut amplitude, $a[\s_c]$,
is obtained from the
momentum factor $a[\s]$ of the Feynman amplitude by replacing the
Feynman propagator $(K^-+i\e)^{-1}$ of every cut line by the
Cutkosky propagator $-2\pi i\d(K^-)$. 
There is then a superficial similarity between a Cutkosky
cut diagram and a nonabelian cut diagram. However, they are completely
different and the cut diagrams referred to in this paper are exclusively
nonabelian cut diagrams.

From \eq{tree}, it follows 
that $a[\s_c]$ is factorized
into products of $a$'s separated by cuts, which is why the formula
is called the {\it factorization} formula. For example, if $[\s_c]=[321|4|5]$,
then $a[\s_c]=a[321]a[4]a[5]$.

We mentioned below eq.~\eq{tree} that the $\d$-function appearing in
the amplitude
must be removed at the end. This can be carried out
by putting in Cutkosky propagators {\it only} where a vertical bar occurs.
The overall $\d$-fucntion for the sum of the `$-$' components of 
momenta then disappears because no cut is ever put after the last entry
$\s_n$ of $[\s]=[\s_1\s_2\cdots\s_n]$.

The {\it complementary cut diagram} (or c-cut diagram for short)
$[\s'_c]$ of a cut diagram $[\s_c]$
is one in which every cut line in $[\s_c]$ becomes uncut, and vice versa.
If no cuts appear in $[\s_c']$,
the colour factor $t[\s_c']$ is simply the product of colour matrices,
as in a Feynman diagram. When cuts
are present, the product straddling an isolated cut is replaced
by its commutator. If consecutive cuts occur, then the product is
replaced by nested multiple commutators. For example,
if $[\s]=[32145]$, then $[\s_c]=
[321|4|5]$, $[\s_c']=[3|2|145]$, and $t[\s_c']=[t_3,[t_2,t_1]]t_4t_5$.

For $n=2$, eq.~\eq{fact} reads $a[12]t[12]+a[21]t[21]=
a[1|2]t[12]+a[21]t[2|1]=a[1]a[2]t_1t_2+a[21][t_2,t_1]$. This
can easily be checked by direct calculation. Explicit
check for $n=3$ is also possible  \cite{LL1},
but for larger $n$ a direct verification becomes very complicated.

The factorization formula \eq{fact} is combinatorial in nature, and is
true whatever the matrices $t_i$ are.
In particular, if all the matrices $t_i$ commute, as in QED, then
the only surviving term on the right-hand side of \eq{fact} is the one
where no commutator occurs, which means that
$[\s_c']$ contains no cuts, and $[\s_c]$ has a cut 
at each propagator. This implies factorization of the tree amplitude
\eq{fact} into a single product $\prod_{i=1}^na[i]$, which  is the usual
eikonal formula \cite{EIK,CW} used in QED scatterings.
It is this factorization that allows exponentiation
to occur and the phase shift to be computed.

In QCD, colour
matrices do not commute,
\be
[t_a,t_b]=if_{abc}t_c,
\labels{cr}
\ee
but their commutators generate other
colour matrices. This allows
the nested multiple commutator of colour matrices 
to be interpreted as a source for a new {\it adjoint-colour} object,
which we call a {\it Reggeon fragment}.
For parton-parton scattering in the weak coupling limit, they turn
out to be the constituents of  the Reggeon as we know it 
from the Regge-pole theory, hence the
name. The word `fragment' is there to clarify that this object is
just part of the Reggeon, not the whole thing, but for brevity
this word is often dropped.
This algebraic characterization of a Reggeon is valid even in the
strong coupling limit, or physical situations other than two-parton
scatterings. Unlike a gluon which is a point-like particle,
a Reggeon is an extended object made up of a bundle of 
gluons, each interacting with the energetic particle at a
different point. Later we may also introduce interactions 
between gluons to tie them together.
Note that there are at least as many distinct Reggeon fragments as there are
nested multiple commutators. With this interpretation,
the cut amplitude $a[\s_c]t[\s_c']$ is nothing but a product of $M$ such
Reggeon amplitudes, with $M-1$ being the number of cuts in $[\s_c]$.

Two final remarks. First, the rule of inserting cuts explained
above depends on how
the gluons are labelled, though at the end of the calculation it
should clearly not matter how it is done. We may adopt the
labelling which is most convenient for our purpose.
Secondly, we have assumed the fast particle to carry a large `+' component
of the light-cone momentum, but we may equally well construct 
cut diagrams if it had a large `$-$' component instead. For parton-parton
scattering, both are required.

\section{Parton-Parton Scattering Amplitudes}
We will study two-body amplitudes  in this section using
cut diagrams.
Cut diagrams are easier to compute than the corresponding
Feynman diagrams, $\ln s$
cancellations occuring in permuted sum of Feynman diagrams 
do not happen in this case, and they are directly related to 
the Reggeon (fragment) amplitudes. These are some of the advantages
of using cut diagrams.

Fig.~2 depicts a QCD
Feynman diagram for the scattering of two energetic
partons. The upper one carries an incoming
momentum
$p_1^\mu=(\sqrt{s},0,\bz)$ in light-cone coordinates, and the lower
one carries an incoming momentum ${p_2}^\mu
=(0,\sqrt{s},\bz)$. Suppose there are $n_1$ gluons attached to the
upper parton and $n_2$ to the lower, then the $n_1!n_2!$ diagrams
obtained by permuting the position of attachment forms a {\it permuted
set} of diagrams. The sum of amplitudes for diagrams in the permuted set
will be called a {\it premuted sum of amplitudes}. 
If we replace the Feynman 
tree attached to each energetic
parton by the corresponding cut tree, as in Sec.~3,
then we have a cut diagram.
According to \eq{fact}, a permuted sum of Feynman amplitudes
is equal to a permuted sum of cut amplitudes, which is why
we may use cut diagrams  instead of
Feynman diagrams for computation. However, when {\it both}
parton trees are permuted we may overcount, in which case the
final result has to be divided by a symmetry factor ${\cal S}$.
This for example will happen to a permuted sum of $s$-channel
ladder diagrams. If $n$ gluons are exchanged between the two
energetic partons, then there are $n!$ distinct diagrams 
by permutation. The permuted set however contains
$(n!)^2$ diagrams, because each distinct
diagram appears $n!$ times in the set.
Hence a symmetry factor ${\cal S}=n!$ is required
for this case.

The scattering amplitude depends on the centre-of-mass energy $\sqrt{s}$
and the coupling constant $g$, as well as 
the momentum transfer $\Delta=|\bd|$,
and the virtuality $Q^2$ if it is involved in deep inelastic scatterings.
Let us concentrate on its
dependence on $g$ and $s$ for the moment. 
The Born amplitude is $g^2s$.
An $\ell$-loop Feynman diagram is proportional to $g^{2(\ell+1)}$, 
but each  loop may also
(though may not) produce a factor of $\ln s$ upon integration. In
this way the amplitude may grow with energy as fast as
$g^2s(g^2\ln s)^\ell$, and there are diagrams doing so.
In the notation of the Introduction, this means that all amplitudes
are bounded by $\sa$. 

Multi-loop Feynman diagrams can usually be computed only in LLA.
When they are summed, their leading power of $\ln s$ unfortunately
cancels in most
$t$-channel colour configurations. See {\it e.g.}
Ref.~\cite{CW} for concrete examples. Sometimes many subleading powers
are cancelled as well. In the case of QED with multi-photon
exchanges, {\it all} $ln s$ powers are cancelled.
This simply means that a non-zero sum can be obtained for these
colour configurations only when individual Feynman diagrams are computed
to the appropriate subleading-log accuracy, which is generally
quite an impossible thing to do.

This disaster is avoided in cut
diagrams \cite{FHL}. This is so because all cancellations to occur
have already taken place in building the individual cut diagrams.
At a cut line,
the Feynman propagator \eq{eik} is replaced by the Cutkosky
propagator $-2\pi i\d(K^-)$. If a $\ln s$ factor is to occur in the
loop involving this Feynman propagator, it comes from
its singularity  at $K^-=0$ through the integral \cite{CW}
\be
\int_{\Lambda^2/s}{dK^-\over K^-+i\e}\sim \ln(s/\Lambda^2),
\labels{ln}
\ee
where $\Lambda^2$ is determined by a mixture of the other scales 
($m^2,\D^2, Q^2$) in the problem, and can be either positive or negative.
In LLA the $\ln\Lambda^2$ factor is dropped so its precise dependence
becomes immaterial.
When this Feynman propagator is replaced by a Cutkosky propagator,
the integral becomes a constant, and the $\ln s$ factor disappears. 
This corresponds to the cancellation of $\ln s$ factors
when Feynman diagrams are
summed, but here the cancellation has already taken place
once the Feynman propagator is changed into the Cutkosky propagator, even
before the high-energy limit is calculated. This is why it is sufficient
to calculate each cut amplitude in LLA.
In contrast, if
we first compute the high-energy limit from individual Feynman diagrams
before summing them,
then cancellation occurs afterwards and each diagram has to
be computed to a subleading-log accuracy.

By symmetry the same thing happens on the lower parton line.
Since a $\ln s$ factor is lost for each cut on the upper parton
line, or the lower (but not both), the $g$ and 
$s$ dependence of a nonabelian cut amplitude with $\ell$ loops,
$m_1-1$ cuts on top, and  $m_2-1$ cuts at the bottom, 
is bounded by 
\be
g^{2M}s(g^2\ln s)^{\ell-M+1},
\labels{lnsm}
\ee 
where $M={\rm max}(m_1,m_2)$.
In other words, it is bounded by $\sa^M$.
Since   $m_1 (m_2)$
is also the number of adjoint-colour Reggeons
emitted from the upper (lower) parton (see
the discussion at the end of Sec.~3),
bounds with different $M$ apply
to different Reggeon (or colour) channels. 
It is this ability of the cut diagram to extract small contributions 
from large-colour channels
that makes it so valuable for unitarization. With this extraction
we can now afford to throw away even smaller contributions
at the same colour. This is what we mean by
the {\it extended leading-log approximation} (eLLA). It differs
from LLA in that it keeps the leading contribution at every
colour, no matter how small  that is.

With eLLA and $t$-channel colour conservation, Reggeon number is
conserved. In other words,
we can ignore diagrams 
with $m_1\not=m_2$. 
This is so because the colour being exchanged must be
contained both in the product of $m_1$ adjoint colours, {\it and}
the product of
$m_2$ adjoint colours. If $m_1>m_2$, say, then there is another diagram,
with $m_1'=m_2$, that possesses all these colours that are allowed,
but with an amplitude
$\sa^{m_2}$ which dominates over the present one whose amplitude is
$\sa^M\sa^{m_1}$. Hence diagrams with $m_1\not=m_2$ can be dropped
in eLLA.

This cut amplitude, with magnitude $\sa^M$ and 
generated by the exchange
of $M$ adjoint-colour objects, is just the
$M$-Reggeon (fragment) amplitude. When all the 
fragments are summed up it becomes the usual $M$-Reggeon
amplitude. In particular, this shows that the most dominant 
comes from a one-Reggeon exchange and is $\sa$, agreeing
with what we know by direct calculation \cite{DEL}.

We shall discuss in the next section the factorization of
Reggeon (fragment) amplitudes.

\section{Factorization, Exponentiation,  and Phase Shifts}
We shall consider in this section
factorization of Reggeon (fragment)
amplitudes into irreducible amplitudes. This factorization 
enables all amplitudes to be summed up into an eikonal form,
and phase shift expressed as the sum
 of all irreducible Reggeon (fragment)
amplitudes.

We will call a Feynman diagram {\it reducible}, if it falls into
$k>1$ disconnected parts after the two energetic-parton lines are
removed. For example, both Fig.~3 and Fig.~4 are both reducible. They have
$k=3$ irreducible parts if each of  a,b,c is irreducible.
A similar definition will be used
for cut diagrams, except that all consecutive uncut gluon lines must be
merged into single Reggeon lines before the two
energetic partons  are removed. For example, Fig.~5 has three irreducible parts
but Fig.~6 has only two. 

Every member in a permuted set of Feynman diagrams has the same number
$k$ of irreducible components, but this is not so in a permuted set
of cut diagrams as evident from the example above. 

In every permuted set there is at least one member diagram which is
{\it uncrossed}. This means the gluon lines from different irreducible
components do not cross one another.
For example, Figs.~3 and 5 are uncrossed diagrams,
and Figs.~4 and 6 are crossed. For reducible diagrams with
$k$ distinct components, there are actually $k!$ uncrossed diagrams obtained
by permuting the relative positions of these components.

To discuss factorization we must first discuss how to construct
and to organize the cut diagrams in a systematic way.
We will do so by induction on the order $n$, assuming that we already 
know how to do it for all 
orders less than $n$.

In this connection the remark made in the last paragraph 
of Sec.~3 about the labelling of gluon lines is relevant. 
The labelling can be assigned arbitrarily for one diagram in
a permuted set, after which the location of cuts in any other
diagram of the set is completely determined.
We shall choose our labellings to  facilitate factorization.

Consider any permuted set of diagrams.  
If $k=1$, there is no question of factorization so we just
label the gluon lines anyway we want to. If $k>1$,
we will choose to fix the labelling from
an uncrossed diagram, such as Figs.~3.
If there are $n_a$
gluons in  part a, $n_b$ in  part b, etc., then label the
gluons in part a by the numbers 1 to $n_a$, and those in part
b by the numbers $n_a+1$ to $n_a+n_b$, etc. This labelling ensures
cuts to occur between any two irreducible components, as in Fig.~5.
The labelling within each irreducible component is assumed to have
been fixed before, and it is correlated with the location of cuts
inside these components. Fig.~5 shows one labelling that gives
rise to the cuts shown.

With cuts between irreducible components, 
the amplitude of the uncrossed cut diagram factorizes in the 
impact-parameter space, 
 \be
\A(s,b)=-2is\prod_j\[ih_j(s,b)\G_j\],
\labels{dg}
\ee
where $\G_j$ is the colour factor of the $j$th irreducible component.
The order of $\G_j$ in the product follows the order of the
irreducible components in the uncrossed diagram.
To see why we have to go to the impact-parameter space for factorization,
and to understand the detailed numerical factors appearing in \eq{dg}, 
let us briefly review the
Feynman rules used to construct the $T$-matrix.
Feynman propagators are taken
to be ${\cal N}/(-q^2+m^2-i\e)$, with ${\cal N}=1, m+\gamma
\.q, -g^{\mu\nu}$
respectively for scalars, spinors, and gauge bosons.
Vertex factors are taken directly from
the coefficients of the interacting Lagrangians. Then all the $i$'s and
$(2\pi)$'s of the T-matrix are contained in the factor 
$[-i/(2\pi)^4]^\ell$, where $\ell$
is the number of loops in the diagram. In other words, we have a factor 
$-i/(2\pi)^4$ per loop. Let $q$ be the loop momentum of any loop containing
a cut on top and a cut
at the bottom. The loop integration for the impact-parameter
amplitude, taken the top and the 
bottom Cutkosky propagators into account, becomes
\be
-i\int {d^2{\bf q}\over (2\pi)^2}e^{i{\bf q}\.\bb}
\h\int {dq^+dq^-\over (2\pi)^2}(-2\pi i)^2\d(\sqrt{s}q^-)
\d(\sqrt{s}q^+)={i\over 2s}\int {d^2\bq\over (2\pi)^2} e^{i\bq\.\bb}.
\labels{ps}
\ee
The factor $2s$ in the denominator of \eq{ps} is cancelled by the
factor $2s$ from a pair of vertices. Since there is one less loop than
pairs of vertices present, an extra factor $2s$ remains, 
which is shown in \eq{dg} in
front of the product sign. This leaves the factor $i$ per loop in \eq{ps},
which accounts for all the $i$ factors in \eq{dg}. Finally, transverse
momentum conservation links the irreducible components together, preventing
factorization to occur. But if we Fourier-transform it into impact-parameter
space, then the exponential factors $\exp(i\bq_i\.\bb)$ separate,
enabling factorization to take place as shown in \eq{dg}.

Once the labelling of gluon lines is fixed this way in the uncrossed
diagram, the location of cuts in any other
cut diagram within the permuted set is completely determined.
A diagram in this set may come from permutation of  lines within
irreducible components, in which case it involves only lower-order
diagrams and by the induction hypothesis we do not have
to worry about it. That leaves diagrams whose gluon lines
from different components cross one another, as in Fig.~6.
Such crossed lines always fuse together some irreducible components
and reduce their total number from $k$ to a number
$k'<k$. If $k'=1$, this is simply a new irreducible component.
If $k'>1$,  then each irreducible component
is of order $<n$, so it must have been included already in 
a lower-order consideration. In any case, the factorization
formula \eq{dg} is still valid, so every cut amplitude
can be factorized into a product of irreducible Reggeon amplitudes.

A permuted set of Feynman diagrams with $k$ irreducible components
has $k!$ uncross diagrams, corresponding to the different
ordering of the components. We could have fixed the labelling 
using any of these. For example, we could fix the labelling
from Fig.~7(a), getting the cuts shown there,
(the thick lines are Reggeon lines,
{\it i.e.,} a group of gluon lines with no cuts between them),
or we could have fixed the labelling from Fig.~7(b). 
The former factorize into something
proportional to $\G_b\G_a\G_c$, and the latter is proportional
to $\G_a\G_b\G_c$. Since the colour matrices do not commute
these two are not the same. Which one should we use? It turns
out that within eLLA it does not matter. Either will do and they
give effectively the same amplitude \eq{dg}, because their
difference is an amplitude that can be neglected in eLLA. This is so
because the difference is proportional to the commutator $[\G_a,\G_b]$.
Using Jacobi identity,
commutator of two nested multiple commutators can be written as sums
of nested multiple commutators. So the difference of Figs.~7(a) and
7(b) is given by cut diagrams with one less Reggeon line, say $M-1$
instead of $M$, which means that the spacetime amplitude $\sa^M$
of these diagrams is negigible compared to the leading contribution
$\sa^{M-1}$ for amplitudes with $M-1$ reggeons.

Summing up all permuted sets of cut diagrams of all orders, we obtain
the complete impact-space amplitude in eLLA to be  
\be
\A(s,b)=2is\[1-\exp\(\sum_j ih_j(s,b)\G_j\)\],
\labels{sum}
\ee
where the sum is taken over all 
irreducible Reggeon fragment amplitudes.  By comparing
this with the formula in \eq{impact}, we deduce immediately the formula
for phase shift to be
\be
2\d(s,b)=\sum_jh_j(s,b)\G_j.
\labels{psf}
\ee

We may combine all the irreducible amplitudes with the same $t$-channel
colour $a$ to write the phase shift as
\be
2\d(s,b)=\sum_ad_a(s,b)\C_a,
\labels{psfa}
\ee
where $\C_a$ is the colour factor for irreducible 
colour $a$. For example,
$\C_8=\gc$ is the adjoint-colour factor,  
where $t'_a$ and $t''_a$ are the colour
matrices of the upper and the lower particles respectively. 
Any diagram of the type Fig.~8(a), for example, will have
a colour factor proportional to $\C_8$ of Fig.~8(b).
We shall also use
$\C_1=\g1$ to denote the singlet colour factor.
Let $M_a$ be the smallest
integer required for irreducible colour $a$ to appear in the product of 
$M_a$ adjoint representations ($M_8=1, M_1=2$, etc.). Then $d_a(s,b)
\sa^{M_a}$, and contribution to this colour from larger number of Reggeons 
can be omitted.

For quark-quark scattering only adjoint and singlet colours can be
exchanged, its phase shift is therefore given by
\be
\d(s,b)=\d_8(s,b)+\d_1(s,b)=d_8(s,b)\gc+d_1(s,b)\g1.
\labels{d81}
\ee 
The scattering amplitude can now be computed from \eq{impact},
and the total cross section from \eq{tot}. To obtain the latter it is
necessary to project out the colour-singlet amplitude $A_1(s,0)$ of $A(s,0)$.
The necessary algebra is carried out in Appendix A, with 
an answer given by eq.~(\ref{a11}). From that one can obtain the total
cross section to be
\be
\s_T(s)&=&2\int d^2\bb\Biggl\{1-{1\over 3}\Biggl[2e^{-2(d_1^I+d_8^I/3)}
\cos\(2d_1^R+2d_8^R/3\)\nn\\
&&\hskip2.7cm +e^{-2(d_1^I-2d_8^I/3)}\cos\(2d_1^R-4d_8^R/3\)\Biggr]\Biggr\},
\labels{totft}
\ee
where $d_i^R$ and $d_i^I$ are respectively the real and imaginary
parts of the  phase shift $d_i(s,b)$.
For gluon-gluon scattering, more colours can be exchanged and more
phase shifts have to be kept.

We shall discuss in the next two sections 
more detailed expressions for the phase shifts in pQCD.

\section{Unitarization of the BFKL Pomeron}
Let $\A_8(s,b)$ and $\A_1(s,b)$ be the adjoint and the singlet components
of the impact-parameter amplitude $\A(s,b)$. Their respective leading
contributions will be denoted by
$\A_8'(s,b)$ and $\A_1'(s,b)$. These can be obtained by
substituting \eq{d81} into \eq{impact} and expanding the exponential.
In this way we obtain
\be 
\A'(s,b)&=&\A_8'(s,b)\gc+\A_1'(s,b)\g1,\nn\\
\A_8'(s,b)&=&-2id_8(s,b),\nn\\
\A_1'(s,b)&=&-2i\(d_1(s,b)+i\xi_2d_8(s,b)^2\),
\labels{ap81}
\ee
where $\xi_2$ is the amount of singlet contained in a pair of adjoints:
\be
t'_at'_b\x t''_at''_b=\xi_2\g1+\cdots.
\labels{ps81}
\ee
For quark-quark scattering, $\xi_2$ is given in (\ref{g1n}) to be
\be
\xi_2=u'u''(u'+u'')={2\over 9}
\labels{xi2}
\ee
for $SU(3_c)$.

Conversely, we can solve for the phase shifts from the leading 
amplitudes to obtain
\be
d_8(s,b)&=&{i\over 2}\A'_8(s,b),\nn\\
d_1(s,b)&=&{i\over 2}\A'_1(s,b)+{i\over 4}\xi_2{A'_8(s,b)}^2.
\labels{psap}
\ee

The impact-parameter amplitude $\A(s,b)$ is related to the momentum-space
amplitude by a Fourier transform. According to \eq{impact}, we have
$\A(s,b)=\bkk{A(s,\D)}/2is$, where
\be
\bkk{F}\equiv{1\over (2\pi)^2}\int d^2\bd e^{-i\bd\.\bb}F(\D).
\labels{bracket}
\ee

It is well known that the leading adjoint amplitude 
$A_8'(s,\Delta)$
is given by the exchange of a Reggeon \cite{DEL}, with
\be
A_8'(s,\Delta)&=&-{2sg^2\over\Delta^2}s^{-\alpha(\Delta)},\nn\\
\alpha(\Delta)&=&{g^2\over 4\pi}N_c\Delta^2I_2(\Delta),
\labels{reggeon}
\ee
where \cite{CW}
\be
I_n(\Delta)\equiv \int\(\prod_{i=1}^n{d^2\bq_i\over(2\pi)^2}
{1\over\bq_i^2+\mu^2}\)(2\pi)^2\d^2\(\sum_{i=1}^n\bq_i-\bd\).
\labels{in}
\ee
The parameter $\mu$ is an infrared cutoff put in by hand.
From \eq{reggeon} we can compute the adjoint phase shift to be
\be
d_8(s,b)&=&{i\over 2}\A_8'(s,b)={1\over 4s}\bkk{A_8'(s,\D)}
=-{g^2\over 2}\bkk{\D^{-2}s^{-\a(\D)}}.
\labels{regge}
\ee
Note that 
\be
I_1(\D)&=&{1\over\D^2+\mu^2},\nn\\
\bkk{I_n}&=&\bkk{I_1}^n.
\labels{i1n}
\ee

Similarly, the 
leading singlet amplitude $A_1'(s,\Delta)$ is given by
the exchange of a BFKL Pomeron \cite{DEL,BFKL,CW,CHENG}.
The details of the Pomeron amplitude is much less well known, even within
LLA. For example, it is known that at extremely high energies,
the amplitude $A_1(s,\Delta=0)$ has an energy dependence of 
\be
s^{1+4\ln 2N_cg^2/4\pi^2},
\labels{bfkl}
\ee
but its complete energy dependence at lower energies is complicated,
even at $\Delta=0$ and within LLA.

\section{Three-Loop Quark-Quark Amplitude}
As discussed in the last section, phase shifts can be
computed from
\eq{d81} and \eq{psap}. Unfortunately the $s$ and $b$
dependences of the BFKL-Pomeron amplitude
are not sufficiently well known to allow us to make a
reliable calculation that way. To get an idea how unitarization
affects the energy dependence of the cross section, and to
illustrate the formalism with concrete formulas, we discuss in this
section the computation of quark-quark scattering phase shift
 to three-loop order.

The quark-quark {\it amplitude} up to two loops
can be found in the book of Cheng and Wu \cite{CW}, where references to the
original literature are given. The three-loop {\it amplitude} can be found in 
Ref.~\cite{CHENG}.

As discussed in \eq{dg} and \eq{psf}, phase shifts can be extracted
from perturbative amplitudes of irreducible colour diagrams. 
To three-loop order, 
one obtains in this way from Table II of Ref.~\cite{CHENG} that
\be
\sum_jh_j(s,b)\G_j&=&h_1\G_1+h_{21}\G_{21}+h_{22}\G_{22}=d_8\gc+d_1\g1,\nn\\
h_1&=&{g^2\over 2}\[-\bkk{I_1}+v\bkk{I_2}-\h v^2\bkk{\D^2I_2^2}+{1\over 6}v^3\bkk{
\D^4I_2^3}\],\nn\\
h_{21}&=&{ig^4\over N_c}\Biggl[v\(\bkk{I_3}-\h\bkk{\D^2I_2^2}\)+v^2\(
\bkk{\D^2I_2I_3}-\bkk{I_1}\bkk{\D^2I_2^2}-\bkk{I_4}\)\nn\\
&+&{2\over 3}v^3\bkk{I_2}\bkk{\D^2I_2^2}\Biggr],\nn\\
h_{22}&=&{ig^4\over N_c^2}v^2\(\h\bkk{\D^4I_2^3}-2\bkk{\D^2I_2I_3}
+\bkk{I_1}\bkk{\D^2I_2^2}+\bkk{I_4}\),\nn\\
v&=&{g^2\over 4\pi}N_c\ln s,
\labels{h}
\ee
where $\G_1,\G_{21},\G_{22}$ are the colour factors for Fig.~8(b), (c), (d)
respectively. 
Strictly speaking, the amplitudes $h_j$ and the colour
factors $\G_j$ are not the ones appearing in \eq{psfa}, because
all colour structures of the form Fig.~8(a) have already been turned
into Fig.~8(b) in the definition of $h_j$ here.  
The function $I_n(\D)$ is defined in \eq{in} and
the brackets $\bkk{F}$ in \eq{bracket}.
The last term in $h_{21}$ is of order $g^{10}$ \cite{CHENG}, 
but it must be included to
keep the exponents in \eq{totft} negative and the integral convergent.

$\G_1$ carries one Reggeon fragment in the $t$-channel and $\G_{12},\G_{22}$
each carries two. We therefore expect $h_1$ to be 
$\sa$ and $h_{21},h_{22}$ to be $\sa^2$. This is precisely what
is shown in \eq{h}.
Table II of Ref.~\cite{CHENG} also contains amplitudes for
the reducible colour factors $\G_1^n$
for $n=2,3,4$, and $\G_1\G_{21}$. It can be verified there that
their corresponding amplitudes can be obtained by the expansion of the
exponential containing the phase shift in \eq{impact}.

Using (\ref{g213}) to obtain the adjoint and the singlet projections 
of $\G_{21}$ and
$\G_{22}$ for $SU(3_c)$, we obtain the phase shifts to be
\be
d_8(s,b)&=&h_1-{1\over 2}h_{21}-{3\over 4}h_{22}, \nn\\
d_1(s,b)&=&{2\over 3}h_{21}+2h_{22}.
\labels{dg8}
\ee

The contribution to the adjoint-colour amplitude can be computed from
\eq{impact}, \eq{ap81}, \eq{h} and \eq{dg8} to be
\be
\A_8'(s,\D)&=&-{2sg^2\over\D^2}\[1-\a(\D)\ln s+{1\over 2!}\(\a(\D)\ln s\)^2
-{1\over 3!}\(\a(\D)\ln s\)^3\]\nn\\
&\equiv& -{2sg^2\over \D^2}\[s^{-\a(\D)}\]_4.
\labels{a88}
\ee
This is just the first four terms of \eq{reggeon} when $e^{-\a(\D)\ln s}$
is expanded. So to order $g^8$ we see directly in this
way that the adjoint amplitude is dominated by the 1-Reggeon exchange.

The leading contribution to the singlet amplitude can be computed
similarly. It is
\be
\A_1'(s,\D)&=&{2isg^4\over 9}\Biggl\{v\(4I_3-2\D^2I_2^2\)+v^2\(
-4\D^2I_2I_3+2\D^4I_2^3\)+{8\over 3}v^3I_2*(\D^2I_2^2)\nn\\
&+&\[{1\over \D^2}s^{-\a(\D)}\]_4*\[{1\over \D^2}
s^{-\a(\D)}\]_4\Biggr\}
\ee
where
\be
F_1(\D)*F_2(\D)\equiv {1\over (2\pi)^2}\int d^2\D'F_1(\D-\D')F_2(\D'),
\labels{conv}
\ee
and an infrared cutoff $\D^2\to\D^2+\mu^2$ should be introduced by
hand  as in \eq{in} and \eq{i1n} to simulate a hadronic size.
This is the BFKL-Pomeron amplitude accurate to order $g^8$.

Since $g_1\equiv h_1, g_{21}\equiv -ih_{21}$,
and $g_{22}\equiv -ih_{22}$ are real, the appropriate
combination used to compute the total cross section from \eq{totft} becomes
\be
2d_1^I+{2\over 3}d_8^I&=&{1\over 2}g_{21}+{7\over 4}g_{22},\nn\\
2d_1^I-{4\over 3}d_8^I&=&g_{21}+{5\over 2}g_{22},\nn\\
2d_1^R+{2\over 3}d_8^R&=&{1\over 3}g_1,\nn\\
2d_1^R-{4\over 3}d_8^R&=&-{2\over 3}g_1.
\labels{dir}
\ee

\section{$\gamma^*P$ and Other Total Cross Sections}
$\gamma^*P$ total cross section will be discussed in this section.
We will concentrate on its energy variation which is
governed by the (unitarized) Pomeron and is thus universal.
The magnitude of the cross section depends on the hadronic size 
so there is no way we can say much about it in pQCD without a model
of the hadron.

The experimental situation is as follows \cite{HERA,ZEUS}.
The variation with energy is consistent with a power growth,
$\s_T(s,Q)\sim s^{a(Q)}$, with a $Q$-dependent exponent $a(Q)$,
which is an increasing function of 
virtuality $Q^2$. Its value at
$Q=0$ is consistent with the universal exponent 0.08 observed in
all hadronic total cross sections \cite{DL}.

Compared to hadronic cross sections, the new and interesting
feature is the dependence of the exponent on $Q$.
We suggest that this may be taken as evidence
that unitarity correction is already
at work at HERA energies, that this dependence on $Q$ is a reflection of
the slowdown of total cross section growth, needed to satisfy 
the Froissart bound $\ln^2s$ asymptotically. 
The fact that the observed exponent is smaller than that calculated from
the LLA BFKL Pomeron lends further
support to this suggestion.

We shall now explain why the $Q$-dependence of the exponent
can be taken as sign of a slowing growth.
To saturate the Froissart bound,
the total cross section asymptotically will have the form
$\s_T(s,Q)=
\s_0(Q)\ln^2(s/\Lambda^2)$, where $\Lambda$
is the scale parameter to measure the energy $\sqrt{s}$ with.
If $Q$ is much larger than the other dimensional
variables in the problem (masses, $\Lambda_{QCD}$, and $\Delta$), 
we expect $\Lambda$ to be determined by $Q$.
On dimensional grounds the simplest dependence would be
 $\Lambda=cQ$ for some dimensionless constant $c$. In any case
$\Lambda(Q)$ is expected to increase with $Q$. Extrapolating
back to HERA energies, the cross section becomes
$\s_T(s,Q)=\s_0(Q)f(s/\Lambda^2(Q))$, for some function
$f(s')$ which approaches
$\ln^2s'$ asymptotically. For a sufficiently small range
of $s$, the energy variation can always be simulated
by a power growth, with the effective exponent
$a=d\ln f(s')/d \ln (s')$ given by the slope of the curve
in a log-log plot. Since this slope is positive at HERA
energies and it must decline to zero at asymptotic energies,
it is reasonable to assume it to be a monotonically decreasing
function of $s'$. In other words, the rate of growth of $f(s')$
decreases with energy.
Now for a given range of $s$, the corresponding range of 
$s'$ becomes smaller for larger $Q$, thus 
placing the data at a region of faster growth.
Hence the effective
exponent $a(Q)$ would be an increasing function of $Q$, as observed.
We can either take that as a prediction for the observed
increase of $a(Q)$ with $Q$, or reverse the argument 
and interpret the observed variation as an indication for the declining
slope of $\ln f(s')$ with increasing $\ln s'$, in an
effort to comply with the Froissart bound at asymptotic energies.

Qualitatively, this prediction of $a(Q)$ increasing with $Q$
 is quite robust, as it is quite independent of the detailed form
of $f(s')$ and $\Lambda(Q)$. If we know the functions $f$ and
$\Lambda$ then the prediction can also be verified quantitatively.
Knowing $\Lambda(Q)$
we can convert the observed $\s_T(s,Q)$  into a function of $s'$
and $Q$. Now plot $\ln\s_T$ against $ln s'$ for every (sufficiently
large) $Q$, and also
$\ln f(s')$ against $\ln s'$. 
The prediction then says that by a suitable up-down movement of
the experimental curves, every one of them can be made to fall on the universal curve $\ln f(s')$
(moving an experimental curve
up and down is equivalent to adjusting the unknown function $\s_0(Q)$).

Are $f(s')$ and $\Lambda(Q)$ calculable in eLLA? 
In principle $f$ is, but unfortunately $\Lambda$ is not.
It is in the nature of a leading-log approximation, extended or not,
that only the highest power term of $\ln s$ is kept. That means even
if we knew $\Lambda$ as in $\ln(s/\Lambda^2)=\ln s-\ln\Lambda^2$,
it would have been dropped during the course of the calculation.
So unless we can go beyond eLLA,
$\Lambda(Q)$ must be taken as a parameter. As to $f(s')$, 
which is the same as $f(s)$ in eLLA, it is given essentially
by \eq{totft} and \eq{psap}. These formulas compute quark-quark
total cross section, not $\gamma^*P$. However, energy variation is
governed by the Pomeron and should be universal, so up to the
unknown function $\s_0(Q)$ we may simply take them to be
the same and take \eq{totft} to be
$f(s)$ for the purpose of the test above.

Even with an unknown $\Lambda(Q)$ the prediction and the test
are still not trivial, because in general
a function $\s_T(s,Q)$ of two variables cannot be fitted by
two functions $\s_0(Q), \Lambda(Q)$ of one variables.

As mentioned before,
the BFKL-Pomeron amplitude is not sufficiently well known as yet
for the universal curve $f(s')$ to be calculated accurately at this
moment. However, we can illustrate this prediction by using
the three-loop result to calculate $f(s')$ approximately \cite{LCONF2}.
A theory of massless
quark-quark scattering lacks an intrinsic distance scale. 
This causes an infrared divergence
which is cut off by the parameter $\mu$ in \eq{in}.
By making a scaling change $\bb\to\mu\bb$, we see from \eq{impact}
and \eq{tot} that  $\s_T(s)=\mu^{-2}\s_T^{(0)}(s)$, with $\s_T^{(0)}(s)$
obtained from $\s_T(s)$ by setting $\mu=1$. Thus  $\mu^{-1}$
is the distance unit to measure the cross sections in.
It simulates hadronic sizes that would come in to provide an infrared
cutoff in hadronic cross sections. It can be considered simply as
a part of the unknown function $\s_0(Q)$. In this way we can use
\eq{psf}, \eq{totft}, and \eq{h} with $\mu=1$
to calculate $\s_T[\gamma^*P]$ and $f(s')$ up to an overall normalization
$\s_0(Q)$.

For the parametric function $\Lambda(Q)$ we shall take a simple
form $cQ+\Lambda_0$. This is simply $cQ$ for large enough $Q$, but
since we do not know where the cutoff of $Q$ is, we put in this
parameter $\Lambda_0$ to accommodate the smaller-$Q$ data.
With $c=4,\Lambda_0=0.2$, and a properly chosen $\s_0(Q)$,  
the computed three-loop result is shown as a solid curve in Fig.~9. 
They agree quite well with the experimental data. In particular,
the $Q$ dependence of the energy growth exponent $a(Q)$ is reproduced.
The dotted
curves show an energy variation of $s^{0.08}$, appropriate for hadronic
total cross sections. They are placed there to show that the rate of energy
growth for the theoretical calculation grows with $Q^2$, and demanded by
the data. The dash curve is $s^{0.5}$, as given by the BFKL Pomeron.

We can use the fitted values of $\s_0(Q), c,$ and $\Lambda_0$
to compute $\sigma(s,Q)/\s_0(Q)$ as a function of $s'=s/\Lambda^2(Q)$.
This is shown in Fig.~10 together with the three-loop energy function
$f(s')$ (solid curve).

\acknowledgements
This research is supported by the National Sciences and Engineering 
Research Council of Canada and the Fonds pour la Formation de Chercheurs et
l'Aide \`a la Recherche of Qu\'ebec.

\appendix

\section{Colour Decomposition}
The necessary colour decompositions for quark-quark scattering
are worked out in this appendix.

The $U(N_c)$ colour matrices $t_\a\ (0\le\a\le N_c^2-1)$
are conventionally normalized to be
\be
\tr(t_\a t_\beta)=\h\d_{\a\beta}.
\labels{trnorm}
\ee
This leads to the completeness relation
\be
(t_\a)_{ij}(t_\a)_{kl}=\h\d_{il}\d_{kj},
\labels{complete}
\ee
where a sum over the repeated index $\a$ is understood.
We shall use the Latin indices, $\a=a\ (1\le a\le N_c^2-1)$,
 to label the $SU(N_c)$
generators. The remaining $U(N_c)$ generator is $t_0=\b1/\sqrt{2N_c}$. 
To simplify writing, we shall use $\bkk{\cdots}$ to denote the trace
$\tr(\cdots)$, and the index $\a$ to represent the generator $t_\a$.
In this notation, the structure constant $if_{\a\beta\gamma}$ 
defined by the commutation
relation
\be
\[t_\a,t_\beta\]=if_{\a\beta\gamma}t_\gamma
\labels{comm}
\ee
is given in terms of the traces of the generators to be
\be
if_{\alpha\beta\gamma}=2\bkk{\a\beta\gamma-\a\gamma\beta}.
\label{fabc}
\ee
It is clear that $f_{\a\beta\gamma}$ is totally antisymmetric in its
indices, and that $f_{0\beta\gamma}=0$.

The following formulas follow from the completeness relation
(\ref{complete}):
\be
\bkk{A\a}\bkk{B\a}&=&\h\bkk{AB},\nn\\
\bkk{Aa}\bkk{Ba}&=&\h\bkk{AB}-{1\over 2N_c}\bkk{A}\bkk{B},\nn\\
\bkk{A\a B\a}&=&\h\bkk{A}\bkk{B},\nn\\
\bkk{Aa Ba}&=&\h\bkk{A}\bkk{B}-{1\over 2N_c}\bkk{AB}.
\labels{ff}
\ee
They will be used to compute colour decompositions.

Let $t'_a$ and $t''_a$ denote the colour matrices of the upper quark
and the lower quark, respectively. A colour diagram like Fig.~8(a)
has the colour factor $G_1\equiv \gc$. The colour factor
for the reducible diagram when this is repeated $n$ times is
\be
G_1^n\equiv t'_{a_1}t'_{a_2}\cdots t'_{a_n}\x  t''_{a_1}t''_{a_2}\cdots 
t''_{a_n}\equiv 
(E_n)_{\mu\nu}t'_\mu\x t''_\nu.
\labels{gn}
\ee
The last expression comes about because the only colour allowed
to be exchanged between two quarks is either a colour singlet
$\g1$, or an adjoint colour $\gc$.
The coefficients $(E_n)_{\mu\nu}$ can be computed from (\ref{trnorm})
to be
\be
(E_n)_{\mu\nu}&=&4\bkk{\mu a_1a_2\cdots a_n}\bkk{\nu a_1a_2\cdots a_n}
\equiv A_n\bkk{\mu\nu}+B_n\bkk{\mu}\bkk{\nu}.
\labels{ab}
\ee
Using (\ref{ff}) to contract the pair of
indices $a_n$, we obtain a recursion relation
for $A_n$ and $B_n$:
\be
A_n&=&\h\(B_{n-1}-{1\over N_c}A_{n-1}\),\nn\\
B_n&=&\h\(A_{n-1}-{1\over N_c}B_{n-1}\).
\labels{recur}
\ee
This pair of equations can be diagonalized using the variables
$C_n^\pm=A_n\pm B_n$, and solved to obtain
\be
C_n^\pm&=&\pm\h\(1\mp{1\over N_c}\)C_{n-1}^\pm=
\pm 4\[\pm\h\(1\mp{1\over N_c}\)\]^n.
\labels{cn}
\ee

Subsituting this into (\ref{gn}) and (\ref{ab}), we obtain
\be
\G_1^n&=&{1\over 4}\({1\over N_c}A_n+B_n\)\g1+\h A_n \gc\nn\\
&=&u'u''\[(u')^{n-1}-(-u'')^{n-1}\]\g1+
\[(u')^{n}-(-u'')^{n}\]\gc,\nn\\
u'&\equiv&\h\(1-{1\over N_c}\),\nn\\
u''&\equiv&\h\(1+{1\over N_c}\).
\labels{g1n}
\ee
From this it follows that
\be
\exp\(2id_8\gc\)&=&\[
u''\exp\(2i u'd_8\)
+u'\exp\(-2iu''d_8\)\]\g1\nn\\
&+&\ \ \[
\exp\(2iu'd_8\)
-\exp\(-2iu''d_8\)\]\gc.
\labels{exp}
\ee
For $SU(3_c)$, $u'=1/3$ and $u''=2/3$, so (\ref{exp}) becomes
\be
\exp\(2id_8\gc\)&=&{1\over 3}\[2e^{2id_8/3}+e^{-4id_8/3}\]\g1
\nn\\
&+&\ \ \[e^{2id_8/3}-e^{-4id_8/3}\]\gc.
\labels{exp3}
\ee
The colour-singlet component of the impact-parameter amplitude
$\A_1(s,b)$ is therefore given by
\be
\A_1(s,b)=1-{1\over 3}\[2e^{2i(d_1+d_8/3)}+e^{2i(d_1-2d_8/3)}\].
\labels{a11}
\ee
Separating the phase shifts into their real and imaginary parts,
$d_i=d_i^R+id_i^I$, and using \eq{tot},
we finally obtain the formula for total cross section in terms of phase
shifts to be
\be
\s_T(s)&=&2\int d^2\bb\Biggl\{1-{1\over 3}\Biggl[2e^{-2(d_1^I+d_8^I/3)}
\cos\(2d_1^R+2d_8^R/3\)\nn\\
&&\hskip2.7cm +e^{-2(d_1^I-2d_8^I/3)}\cos\(2d_1^R-4d_8^R/3\)\Biggr]\Biggr\}.
\labels{totf}
\ee

For the purpose of calculating phase shifts from perturbation
theory, other colour decompositions are required, especially
those with two $t$-gluon lines in the colour diagram. Let $\G_{2n}$
be the colour factor for a colour diagram with 2 $t$-gluons and $n$
`horizontal' gluons. For example, the colour factor for Fig.~8(c) is 
$\G_{21}$ and the colour factor for Fig.~8(d) is $\G_{22}$. The colour factor
of a triple gluon vertex in these diagrams is $if_{abc}$, 
read counter-clockwise, so that 
\be
\G_{21}&=&(if_{ace})(if_{edb})t'_at'_b\x t''_ct''_d,\nn\\
\G_{22}&=&(if_{ahg})(if_{ceh})(if_{fed})(if_{fbg})t'_at'_b\x t''_ct''_d.
\labels{g22d}
\ee
We will now show that
\be
\G_{2n}&=&{1\over 4}N_c^{n-2}\(N_c^2-1\)\g1-\h\({N_c\over 2}\)^{n-1}\gc.
\labels{g2n}
\ee
In particular, for $SU(3_c)$,
\be
\G_{21}&=&{2\over 3}\g1-\h\gc,\nn\\
\G_{22}&=&2\g1-{3\over 4}\gc.
\labels{g213}
\ee

Eq.~(\ref{g2n}) can again be obtained by induction, as follows. First of
all, define $I_n(\alpha\beta|\gamma\delta)$ by
\be
\G_{2n}=I_n(\a\beta|\gamma\delta)t'_\a t'_\beta \x t''_\gamma t''_\delta,
\labels{i}
\ee
where $I_n$ contains $n$ pairs of triple-gluon colour factors $(if_{\cdot
\cdot\.})$.
See Figs.~6(c),(d),(e). Strictly speaking, the subscripts of $t'$ and $t''$
should be Latin indices and not Greek, but since $f_{0\.\.}=0$,
we may extend the subscripts to Greek indices as shown in (\ref{i})
for $n\ge 1$. For $n=0$, we should identify $\G_{20}$ with $\G_1$.
Next, as only singlet and adjoint colours can be exchanged, we can
write
\be
\G_{2n}&\equiv&(F_n)_{\mu\nu}t'_\mu\x t''_\nu\equiv P_n\g1+ Q_n\gc ,\nn\\
(F_n)_{\mu\nu}&=&4\bkk{\mu\a\beta}I_n(\alpha\beta|\gamma\delta)
\bkk{\nu\gamma\delta}.
\labels{gfi}
\ee
Now write a recursion formula for $I_n(\alpha\beta|\gamma\delta)
\bkk{\nu\gamma\delta}$, as indicated in Fig.~8(f):
\be
I_n(\alpha\beta|\gamma\delta)\bkk{\nu\gamma\delta}&=&
I_{n-1}(\alpha\beta|\rho\s)(if_{\gamma\tau\rho})(if_{\sigma\tau\delta})
\bkk{\nu\gamma\delta}\nn\\
&=&I_{n-1}(\alpha\beta|\rho\s)
\bkk{\nu[\tau,\rho][\s,\tau]}.
\labels{irecur}
\ee
This is valid for all $n\ge 1$, though we should note that $\G_{20}=\G_1^2$,
hence $I_0(\a\beta|\rho\s)$ vanishes unless all four are $SU(N_c)$
indices, in which case $I_0(ab|rs)=\d_{ar}\d_{bs}$.
The repeated $\tau$ index can be summed up using (\ref{ff}) to get
\be
\bkk{\nu[\tau,\rho][\s,\tau]}&=&\h\(-\bkk{\rho}
\bkk{\nu\s}-\bkk{\s}\bkk{\nu\rho}+\bkk{\nu}\bkk{\rho\s}+\bkk{1}\bkk{\nu\rho\s}\).
\labels{recursum}
\ee
The first two terms do not vanish only for $\nu=0$ and $\sigma=0$,
but then since
$f_{0\cdot\cdot}=0$, these two terms never contribute.
This leaves the last two terms.
It is now convenient to consider separately $\nu=0$ and $\nu=n$.
In the first case, the last two terms are identical and the
right hand side of (\ref{recursum}) becomes $\sqrt{N_c/2}\bkk{\rho\sigma}
=\sqrt{N_c/8}\d_{\rho\s}$. Now
$I_{n-1}(\a\beta|\rho\s)$ contains $(n-1)$ pairs of triple-gluon
colour factors. Its bottom pair takes on the form $(if_{\rho\phi\zeta})
(if_{\xi\phi\sigma})$. When contracted with $\d_{\rho\s}$,
we get $(if_{\rho\phi\zeta})(if_{\xi\phi\rho})=N_c\d_{\zeta\xi}$. Hence
\be
I_{n-1}(\alpha\beta|\rho\s)\d_{\rho\sigma}&=&N_c
I_{n-2}(\alpha\beta|\zeta\xi)\d_{\zeta\xi}=\cdots
=N_c^{n-1}I_0(\a\beta|\rho\s)\d_{\rho\s}.
\labels{ire}
\ee
Substituting this back into eqs.~(\ref{irecur}) and (\ref{gfi}),
one gets
\be
(F_n)_{\mu 0}&=&4\bkk{\mu\a\beta}\sqrt{N_c/8}N_c^{n-1}
I_0(\alpha\beta|\rho\s)\d_{\rho\s}\nn\\
&=&\sqrt{2N_c}N_c^{n-1}\bkk{\mu aa}=\h\d_{\mu 0}N_c^{n+1}\(1-{1\over N_c^2}\),\nn\\
P_n&=&{1\over 2N_c}F_{00}={1\over 4}N_c^n\(1-{1\over N_c^2}\).
\labels{p}
\ee
This expression of $P_n$ agrees with the one quoted in (\ref{g2n}).

Now we return to (\ref{recursum}) and consider the case when
$\nu=n$. In that case the right hand side of that equation becomes
$N_c\bkk{n\rho\sigma}$, so we obtain a recursion relation from (\ref{gfi})
to be
\be
(F_n)_{\mu n}&=&\h N_c(F_{n-1})_{\mu n}.
\labels{frecur}
\ee
Since the adjoint component of $\G_{20}=\G_1^2$ is given in (\ref{g1n})
to be $Q_0=(u')^2-(-u'')^2=-1/N_c$, we can solve (\ref{frecur})
to obtain
\be
Q_n=\h N_cQ_{n-1}=\({N_c\over 2}\)^nQ_0=
-\h \({N_c\over 2}\)^{n-1},\labels{q}
\ee
agreeing with the expression given before. This completes the
proof of the formula quoted before in (\ref{g2n}).

\newpage

\begin{figure}[h]
\vspace*{5cm}
\includegraphics{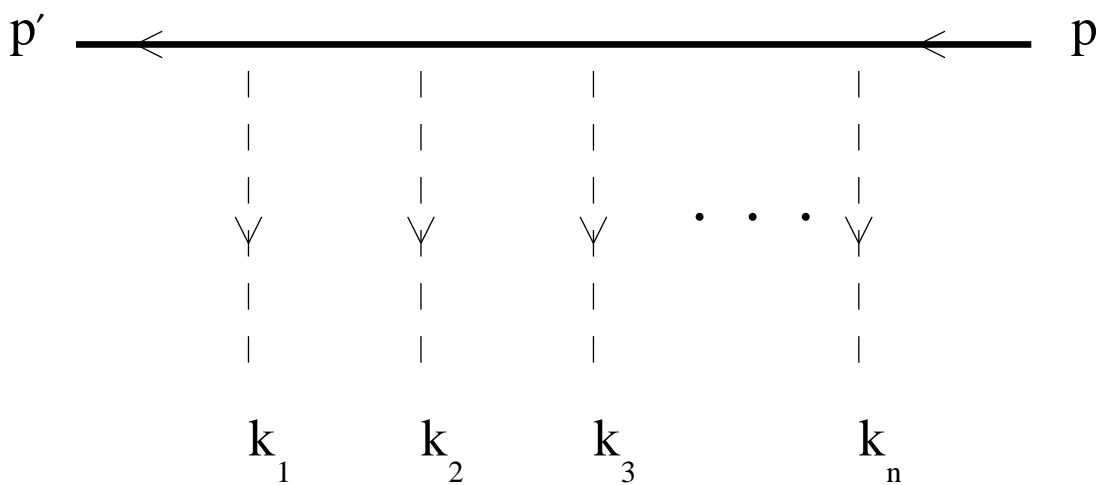}
\vspace*{2cm}
\caption[]{A tree diagram depicting $n$ quanta of 
momenta $k_i$
emitted from an energetic particle.}
\end{figure}

\begin{figure}[h]
\vspace*{10cm}
\includegraphics{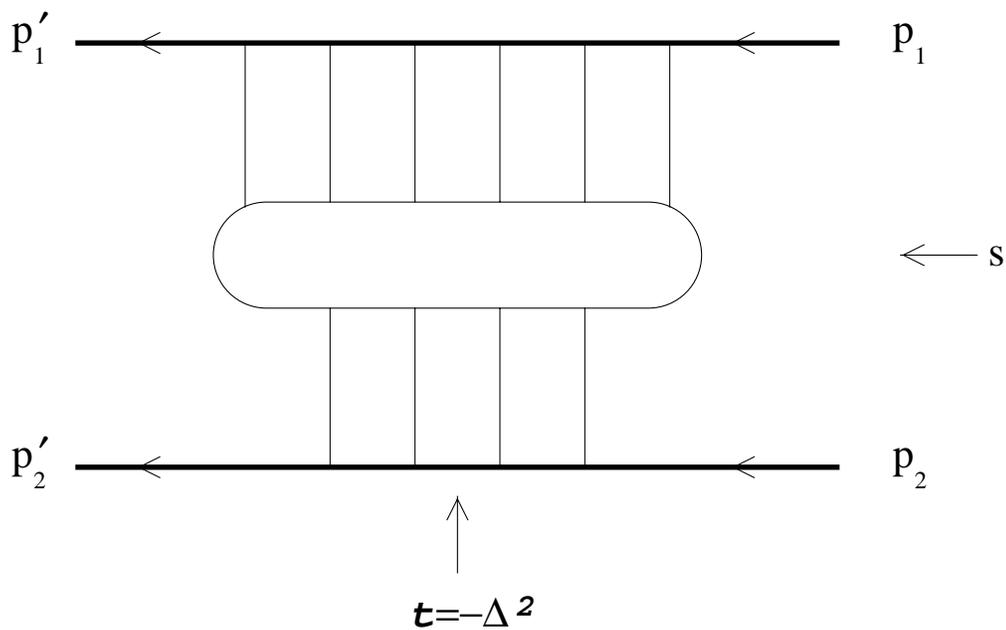}
\vspace*{1cm}
\caption[]{A high-energy scattering diagram with cm energy $\sqrt{s}$
and momentum transfer $\Delta$.}
\end{figure}

\begin{figure}[h]
\vspace*{5cm}
\includegraphics{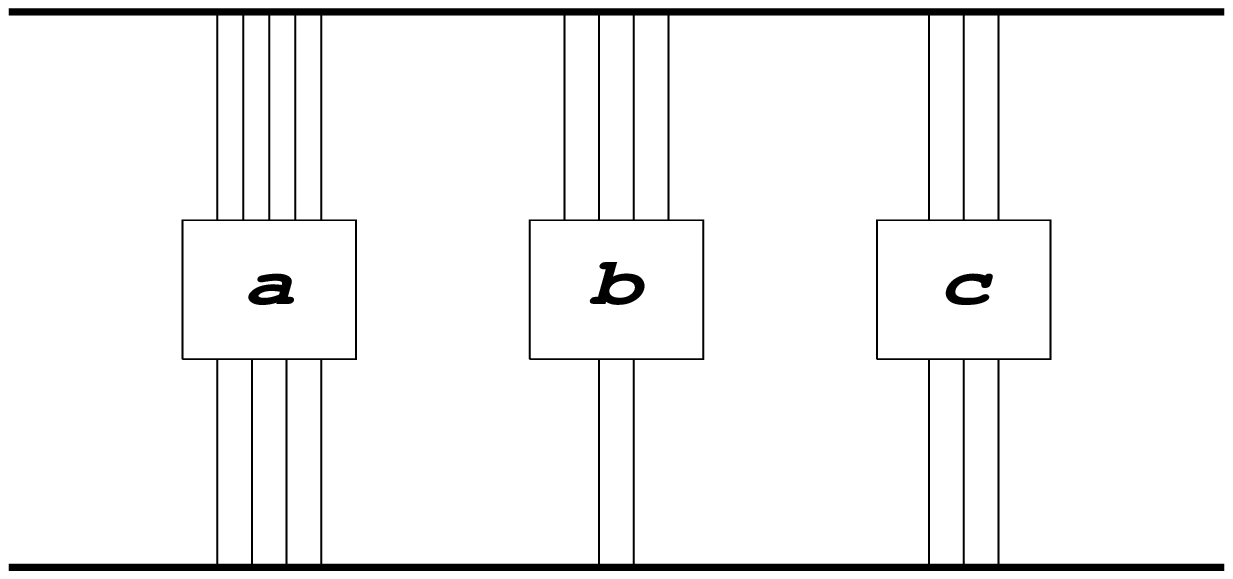}
\vspace*{1cm}
\caption[]{An uncrossed Feynman diagram with three irreducible components.}
\end{figure}

\begin{figure}[h]
\vspace*{9cm}
\includegraphics{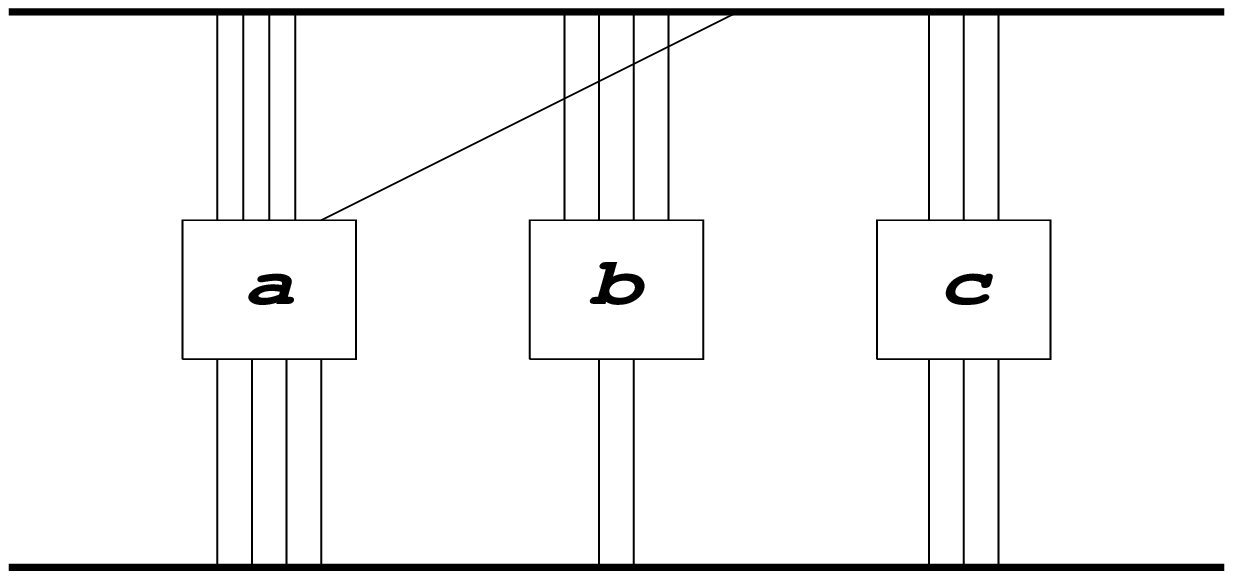}
\vspace*{1cm}
\caption[]{A crossed Feynman diagram with three irreducible components.}
\end{figure}

\begin{figure}[h]
\vspace*{5cm}
\includegraphics{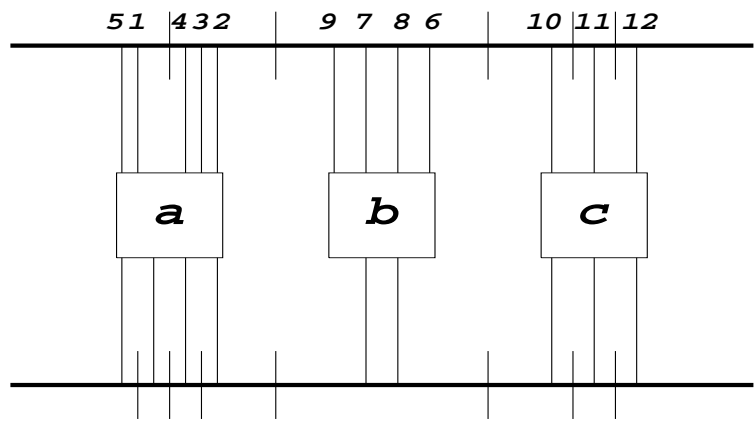}
\vspace*{1cm}
\caption[]{An uncrossed cut diagram with three irreducible components.}
\end{figure}

\begin{figure}[h]
\vspace*{9cm}
\includegraphics{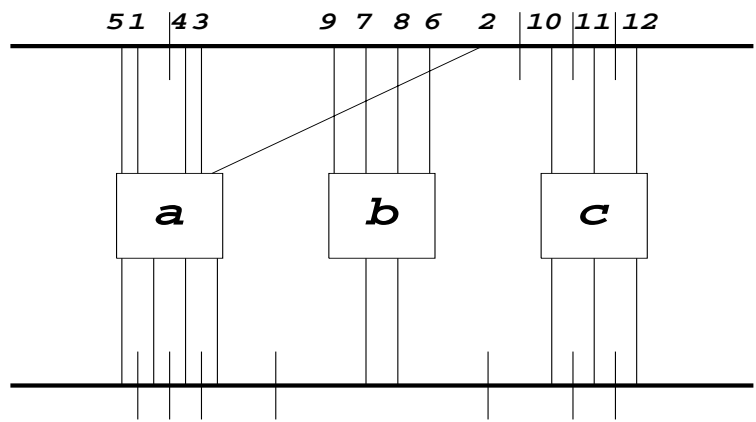}
\vspace*{1cm}
\caption[]{A crossed cut diagram with two irreducible components.}
\end{figure}

\vskip5cm

\begin{figure}[h]
\vspace*{12cm}
\includegraphics{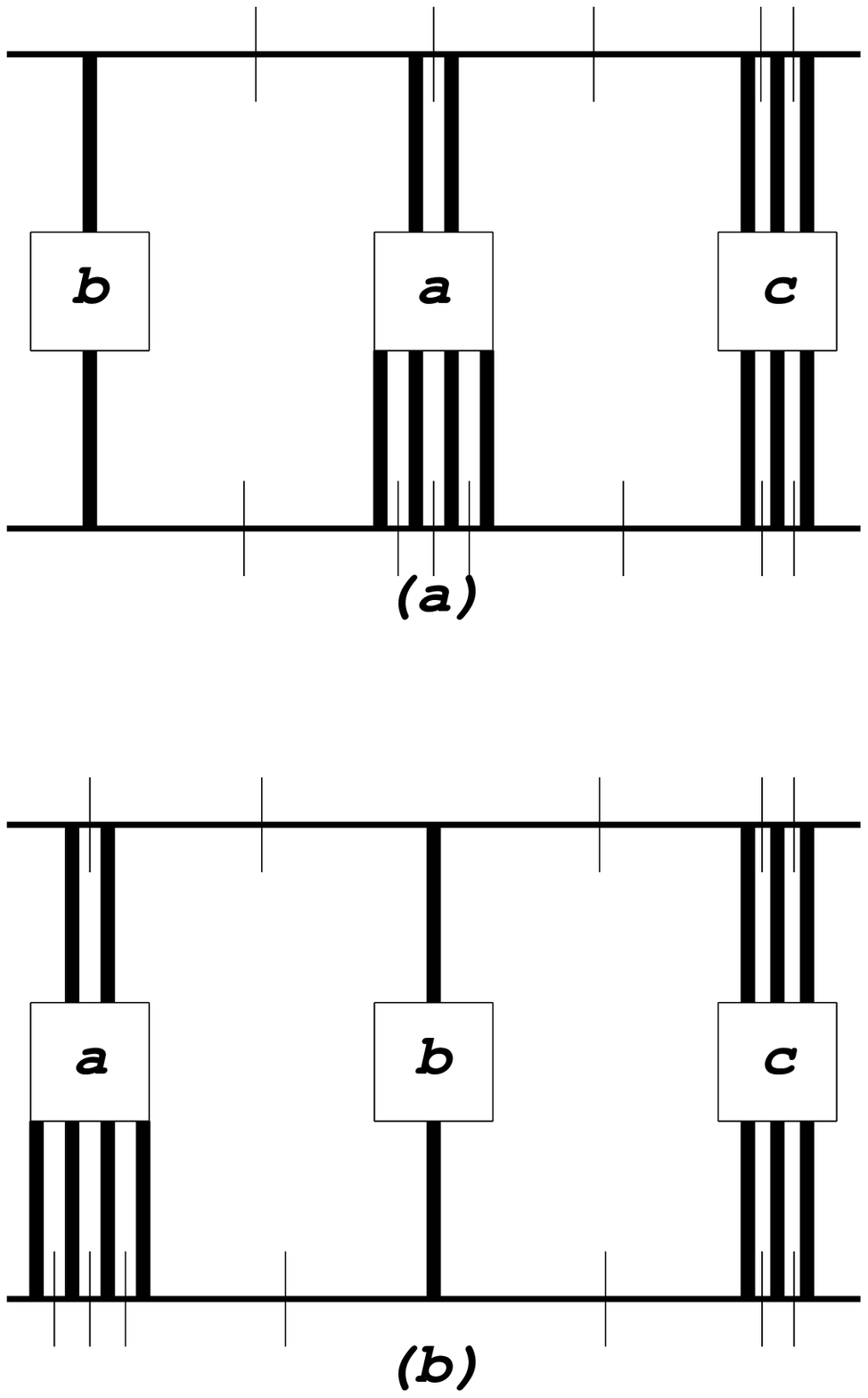}
\vspace*{6cm}
\caption[]{Two uncrossed cut diagrams differing from each
other by the permutation of the first two components. 
The thick lines are Reggeon fragments.}
\end{figure}

\begin{figure}[h]
\vspace*{12cm}
\includegraphics{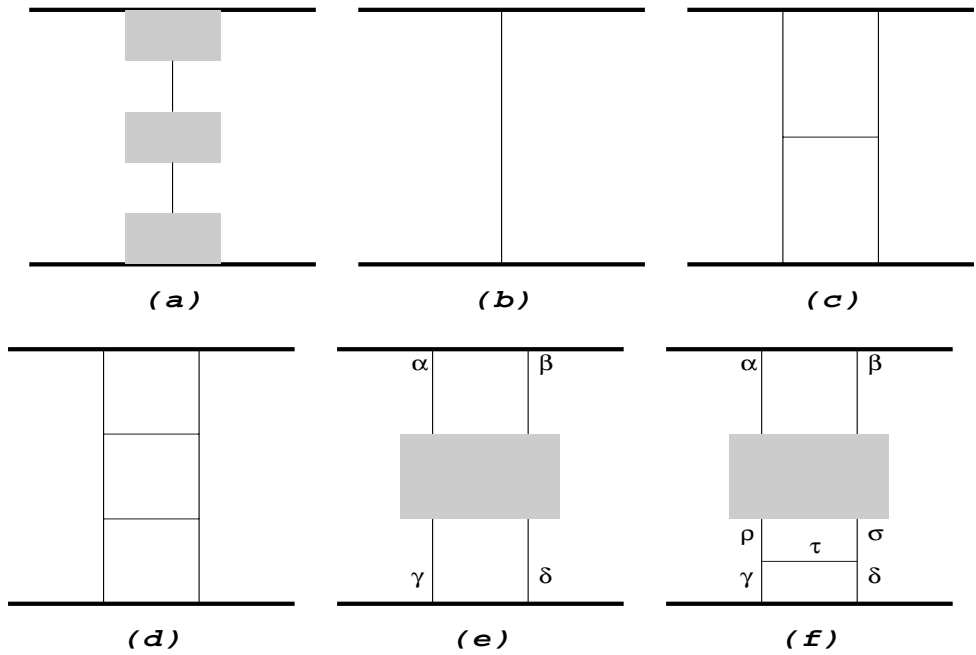}
\vspace*{3cm}
\caption[]{Various colour diagrams referred to in the text and in
Appendix A.}
\end{figure}

\begin{figure}[h]
\vspace*{12cm}
\includegraphics{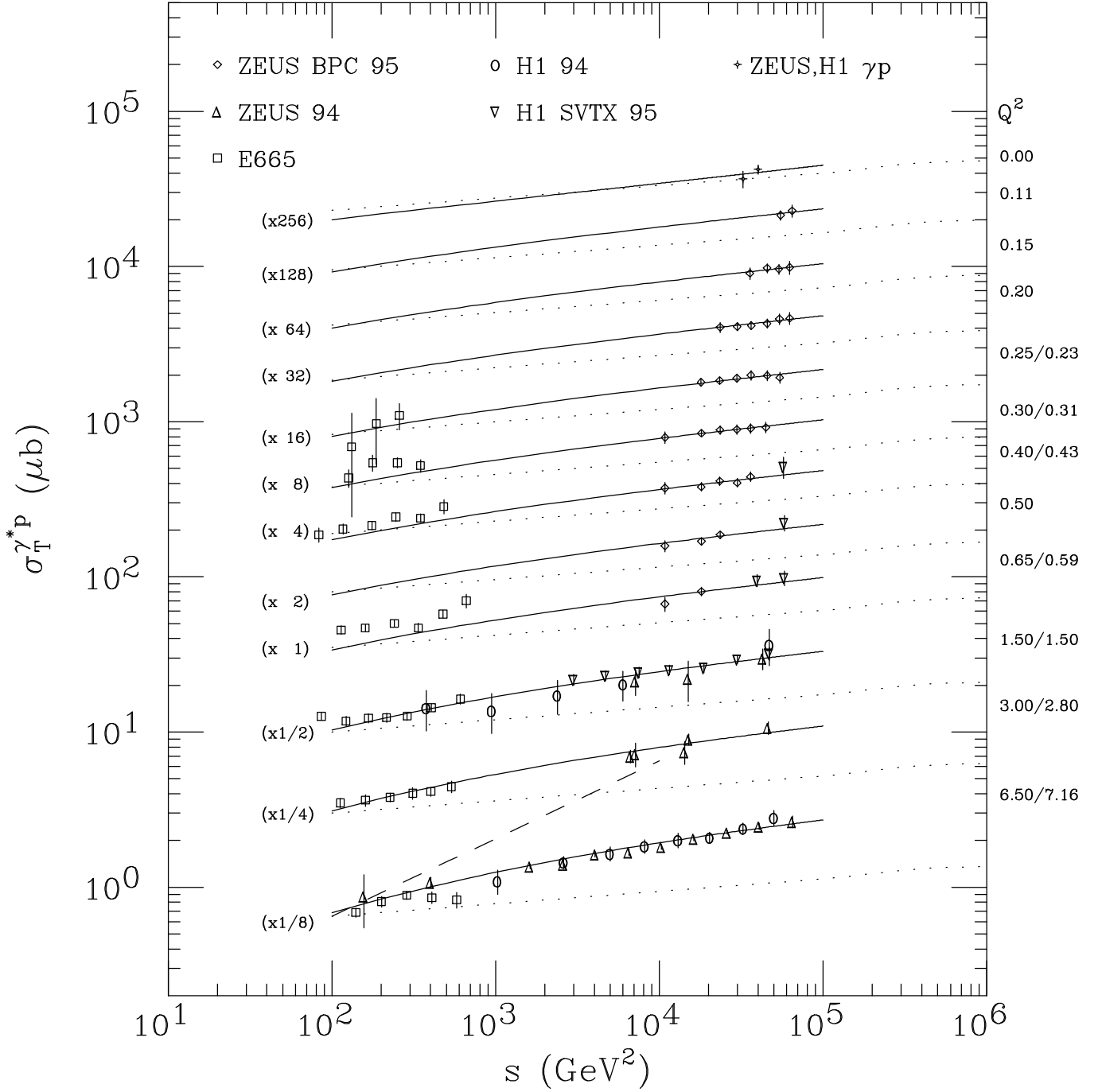}
\vspace*{7.5cm}
\caption[]{$\gamma^*P$ total cross section as a function of photon
virtuality $Q^2$ and c.m. energy $\sqrt{s}$. The solid curve 
is a fit of the three-loop eq.~\eq{h}, with $\Lambda(Q)=[0.2+4Q]GeV$ and 
$\mu(Q)$ chosen arbitrarily. The dotted and the dashed curves represent
an $s^{0.08}$ and an $s^{0.5}$ variation respectively.
Data are taken from Ref.~\cite{ZEUS}.}
\end{figure}

\begin{figure}[h]
\vspace*{12cm}
\includegraphics{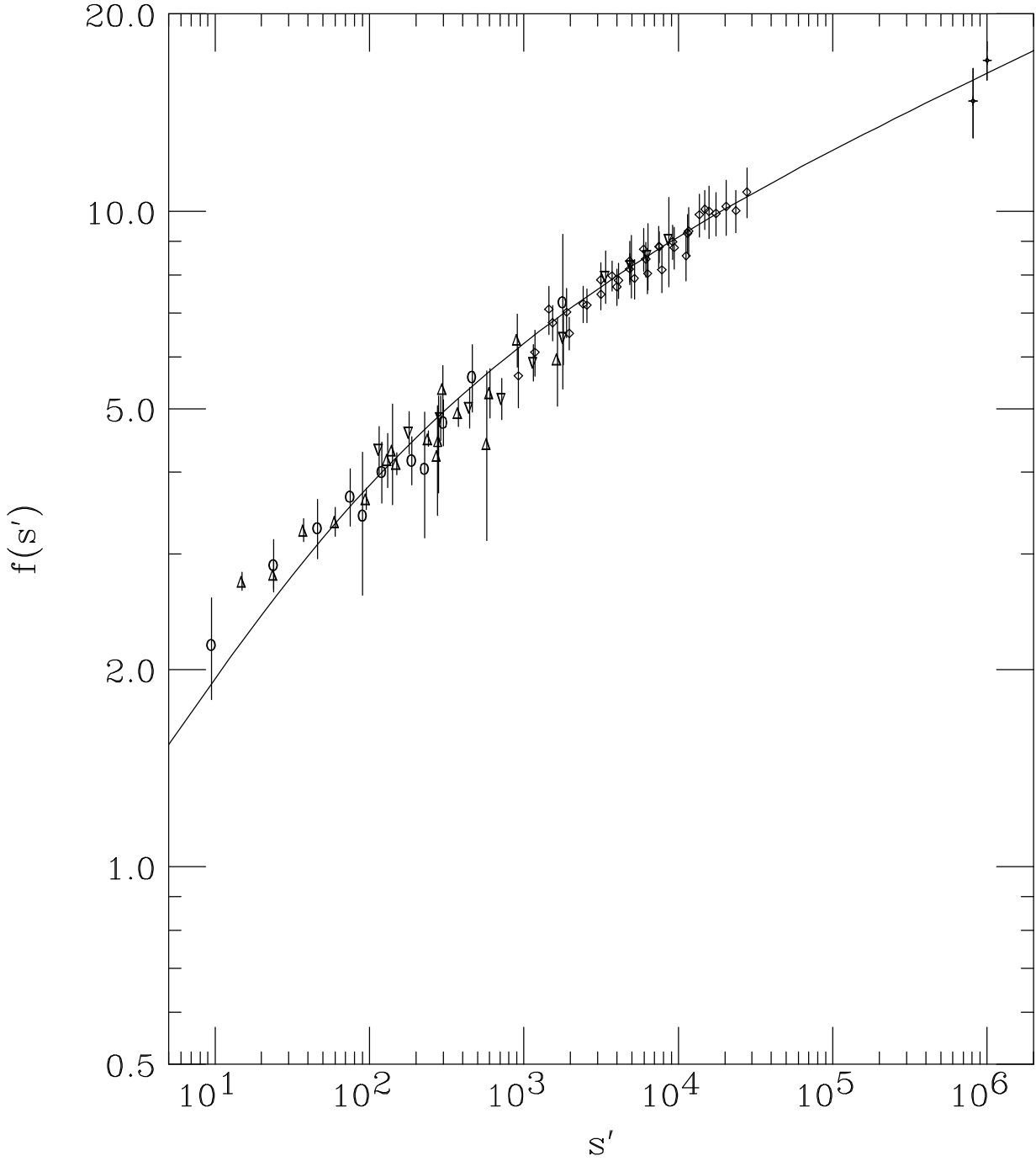}
\vspace*{7.5cm}
\caption[]{The universal energy variation curve $f(s')$ obtained
from the three-loop formula \eq{h}, with arbitrary normalizations.
The data are those of Fig.~6, with the same $\Lambda(Q)$, but
with all low energy points with $s<1000(GeV)^2$ removed.}
\end{figure}

\end{document}